\documentclass[aps,prab,groupedaddress,floats,reprint]{revtex4-2}

\usepackage[english] {babel}
\usepackage{amsfonts,graphicx,soul,natbib,mathrsfs}

\usepackage[colorlinks=true]{hyperref}

\usepackage{epstopdf}
\usepackage{xcolor}
\usepackage{soul}
\usepackage{physics}
\usepackage{graphics}
\usepackage{comment}
\newcommand{\p}{\partial}
\renewcommand{\vec}[1]{\textnormal{\boldmath$#1$}}

\usepackage{dcolumn}
\usepackage{bm}
\begin{document}

\title{Flat bubble regime and laminar plasma flow in a plasma wake field accelerator.}

\author{S. S. Baturin}%
\email{s.s.baturin@gmail.com}%
\affiliation{School of Physics and Engineering,
ITMO University, 197101 St. Petersburg, Russia}%

\date{\today}
\begin{abstract}
   A simple 2D model of the bubble formation in a plasma wakefield accelerator is developed and investigated. It is shown that in the case of a flat driver the bubble may consist of two parts that correspond to two different types of the plasma flow: a laminar flow where  plasma electron streams do not cross and a two-stream (turbulent) flow. The laminar flow turns out to be robust to the symmetry breaking. Building-of of the developed model we demonstrate that in the case of the laminar flow and non-relativistic plasma electrons the transverse wake field is absent inside the bubble even in the case of a transversely nonuniform plasma.       
\end{abstract}

\maketitle
\section{Introduction}

Plasma wakefield acceleration (PWA) technique \cite{GVoss,Chen,Rosen1,Rosen2} attracted many attention over the past several years and is envisioned as one of the main options for the design of future collides and light sources \cite{SNM3,LK,alegro-2019-a,Adli,Sasha1,Jamie1}. 

Development of the PWA emerged into the rapid development of PWA theories from simple yet powerful analytical models \cite {Barov,Lu,Stupakov:2016,Stupakov:2018,Mori} to complex and precise simulation codes \cite{QPIC,warp,OSIRIS}. Nonetheless, significant understanding of physics processes that underlies PWA technology has already been gained there is still an ongoing work on the a theoretical side as well as extensive experimental developments. Several question such as optimal beam-loading \cite{BL}, instability suppression and instability mechanisms \cite{Timon1,Timon2,Lehe} as well as acceptable tolerances \cite{Carl} are still under active study.

One of the main difficulties that is common to all wakefield accelerators be it a plasma based or a structure based (dielectric loaded or corrugated structures \cite{Jing,Sasha2,Sasha3}) machines is the beam beak-up instability that directly affects the efficiency of the accelerator \cite{Hosing1,BBU2,BBU3}. Many mechanisms and approaches including ion motion \cite{Timon2} BNS damping \cite{Lehe,Timon1} and other methods of resonance illumination has been investigated.    

One of the approaches that stands aside is the transverse shaping of the driver beam that mainly includes injection of the flat beam (or the beam with the high ellipticity) \cite{Piot1,Piot2,Piot2c} and dual driver injection \cite{Baturin:dDrive}. It was predicted \cite{FLb,Baturin:dDrive} and recently experimentally demonstrated \cite{Brendan:2020} that highly elliptic driver can suppress transverse wake in a structure based wakefield accelerating device.
It turns out that asymmetric drivers are favorable and promising approach in the hollow channel plasma as well \cite{Joshi}. 

It has been recently realized that path towards a real machine unavoidably includes staging of the individual cells and this in turn results in tolerances in plasma cell alignments as well as on tolerances in plasma uniformity. The indirect evidence of the non-uniformity effects could be seen from the pioneering experiment on hollow channel plasma described in Refs.\cite{Gessner2,Gessner1}. 

Being motivated by the recent success on the flat beam application and the need of preliminary analysis on the transverse plasma non-uniformity, in this paper we develop a simple analytical model of the blow out plasma regime formed by the infinitely flat driver and investigate how local non-uniformity of the plasma density affects the problem. We demonstrate that in contrast to the similar cylindricaly symmetric problem with the point driver considered in Refs.\cite{Stupakov:2016,Stupakov:2018} in the planar case the plasma has two regimes, namely a laminar flow and a turbulent flow. We demonstrate as well that in the case of a laminar flow and non-relativistic plasma electrons the transverse wake is absent even in the case of the transversely nonuniform plasma. 

The model and results of the presented analysis may serve as a starting point for further investigation of the flat beam injection into the nonuniform plasma as well as may be useful for some basic parameter estimations for the ongoing experimental effort at the Argonne Wakefield Accelerator facility \cite{Gerard}.

\section{Basic equations}

In this section we provide a brief overview of the the model that we utilize for the calculus and analysis.

\subsection{General formulas}

We start from the set of equations derived in Ref.\cite{Stupakov:2018}.  We use the same convention as the Ref.\cite{Stupakov:2018} and we use dimensionless variables: time is normalized to $\omega_p^{-1}$, length to $k_p^{-1}$, 
velocities to the speed of light $c$, and momenta to $mc$. We also normalize fields to $mc\omega_p/e$, forces to $mc\omega_p$, potentials to $mc^2/e$, the charge density to $n_0 e$, the plasma density to $n_0$, and the current density to $en_0c$. With $e$ being the elementary charge, $e>0$.

Equation of motion for the plasma electrons could be written as 
\begin{align}
\label{eq:ple}
\frac{d\mathbf{p}}{dt}=\mathbf{\nabla} \psi+\mathbf{\hat{z}}\cross \mathbf{B}_\perp-\mathbf{v}\cross\mathbf{B},~\frac{d\mathbf{r}}{dt}=\frac{\mathbf{p}}{\gamma}.
\end{align}
Here $\mathbf{p}$ is the momentum of the plasma electrons, $\gamma=\sqrt{1+p^2}$ is the relativistic gamma factor of the plasma electrons, $\mathbf{v}=\mathbf{p}/\gamma$ is the velocity and $\psi=\phi-A_z$ is the pseudo potential that defines the wake field as
\begin{align}
\label{eq:lor_main}
E_z=\frac{\partial \psi}{\partial \xi}, ~~~\mathbf{F}_{\perp}=-\mathbf{\nabla}_{\perp}\psi,
\end{align}
and $\mathbf{\nabla}=(\partial_x,\partial_y,-\partial_\xi)$.
Here $\mathbf{F}_{\perp}$ is the transverse part of the Lorentz force  per unit charge of the test particle and $\xi=t-z$. 

Eqs.\eqref{eq:ple} have the following integral of motion
\begin{align}
\gamma-p_z-\psi=1,
\end{align}
as a consequence we have 
\begin{align}
1-v_z=\frac{1+\psi}{\gamma}.
\end{align}
In a quasi-static picture it is convenient to replace the derivative by time $t$ with the derivative by $\xi$.
We use the fact that
\begin{align}
\frac{d\xi}{dt}=1-v_z, 
\end{align}
consequently for an arbitrary function $f(\xi)$ we have.
\begin{align}
\label{eq:con}
\frac{df}{dt}=\frac{df}{d\xi}\frac{d\xi}{dt}=(1-v_z)\frac{df}{d\xi}=\frac{1+\psi}{\gamma} \frac{df}{d\xi}.
\end{align}
Since in a quasi-static picture momenta of the plasma electron is a function of $\xi$ Eqs.\eqref{eq:ple} with Eq.\eqref{eq:con} are reduced to
\begin{align}
\frac{d\mathbf{p}_\perp}{d\xi}=\frac{\gamma}{1+\psi}\mathbf{\nabla}_\perp \psi+\mathbf{\hat{z}}\cross \mathbf{B}_\perp-\frac{B_z}{1+\psi}\mathbf{p}_\perp\cross\mathbf{\hat{z}}.
\end{align}
Equation for the pseudo potential reads
\begin{align}
\label{eq:WP}
\Delta_\perp\psi=(1-v_z)n_e-n_i(x),
\end{align}
 here $n_e$ is the plasma electron density and $n_i(x)$ is the ion density that depends on $x$. In what follows, we will assume that
 \begin{align}
     n_i(x)
     =
     1+gx,
 \end{align}
 with $g\ll 1$.
 Equations for the magnetic field are 
 \begin{align}
& \Delta_\perp B_z=\mathbf{\hat{z}}\cdot(\mathbf{\nabla_\perp}\cross n_e\mathbf{v}_\perp), \\
 &\Delta_\perp \mathbf{B}_\perp=-\mathbf{\hat{z}}\cross\mathbf{\nabla}_\perp n_e v_z-\mathbf{\hat{z}}\cross \partial_\xi n_e \mathbf{v}_\perp.
 \end{align}

The final equation that complements equations for the fields is the continuity equation for the plasma given by  
\begin{align}\label{eq:cont_equation}
    \p_\xi[n_e(1-v_z)]
    +
    \nabla_\perp
    \cdot
    n_e\vec v_\perp
    =
    0
    .
\end{align}

\subsection{2D approximation}

If one assumes that the charge density is uniform and the charge is distributed along the infinite line in $y$ - direction then the following simplification to the main set of equations could be made. Eq.\eqref{eq:lor_main} reduces to
\begin{align}
\label{eq:2dF}
    E_z=\frac{\partial \psi}{\partial \xi}, ~~~F_x=-\frac{d\psi}{dx}, ~~~F_y=0.
\end{align}
Equation for the pseudo potential \eqref{eq:WP} reads

\begin{align}
\label{eq:2dpp}
    \frac{d^2 \psi}{dx^2}=(1-v_z)n_e-n_i,
\end{align}

and continuity equation \eqref{eq:cont_equation} reads

\begin{align}
\label{eq:cont2d}
    \p_\xi[n_e(1-v_z)]
    +
    \frac{ d}{dx} n_e v_x
    =
    0
    .
\end{align}

\section{Shock wave}

As discussed in details in Ref.\cite{Stupakov:2018} plasma electrons cross an infinitesimally thin layer in which the fields produced by the driver 
have a delta-function discontinuity

\begin{align}
    \mathbf{E}_\perp&=\mathbf{D} \delta(\xi), \nonumber \\
    \mathbf{B}_\perp&=\mathbf{\hat{z}}\times \mathbf{D} \delta(\xi) 
\end{align}

with the transverse profile defined by the vector $\mathbf{D}=(D_x,D_y)$.

To solve for the shock wave at $\xi=0$, we assume that the plasma density in front of the moving driver has a linear gradient
    \begin{align}\label{14}
        n_e^{(0)}
        =
        1 + gx
        ,
    \end{align}
where the uniform part of the density in the units that are introduced is 1, $g$ is a constant, and $x$ is the transverse coordinate. We will assume a small gradient,
    \begin{align}
        g\ll 1
        ,
    \end{align}
and use the perturbation theory.

We consider equation for the vector $\mathbf{D}$ that according to Ref.\cite{Stupakov:2018} reads
\begin{align}
\label{eq:Dg}
\Delta_\perp \mathbf D =\frac{n_e^{(0)}}{\gamma_0} \mathbf{D}.
\end{align}
If we set $g=0$ then due to the symmetry we have $D_y^{(0)}=0$ and $D_x^{(0)}=D_x^{(0)}(x,\xi)$.
This immediately results in the equation for $D_x^{(0)}$ in a form
\begin{align}
\label{eq:Dr2}
&\frac{d^2}{dx^2}D_x^{(0)} =\frac{n_e^{(0)}}{\gamma_0} D_x^{(0)}.
\end{align}
With the unmodified plasma density and initial gamma set to the unity ($n_e^{(0)}=1$, $\gamma_0=1$) for boundary conditions $D_x(\pm \infty)=0$, $D_x(x\to \pm 0)=\pm 2\pi \lambda$, where $\lambda$ is the line charge density of the beam, we have
\begin{align}
\label{eq:D01}
    D_x^{(0)}=\left \{ \begin{array}{c}
 2\pi \lambda \exp(-x)~~x>0 \\
 -2\pi \lambda \exp(x)~~~x<0 \\
\end{array} \right. .
\end{align}
Now we consider $n_0$ as given by Eq.\eqref{14}. We apply perturbation theory and introduce anzats
\begin{align}
\label{eq:ppr221}
    &D_x=D_x^{(0)}+g D_x^{(1)}, \\
    &D_y=D_y^{(0)}+g D_y^{(1)}. \nonumber
\end{align}

Substituting Eqs.\eqref{eq:ppr221} and Eq.\eqref{14} into the Eqs.\eqref{eq:Dg}, equating terms of the same order and assuming for simplicity $x>0$ we arrive at a set of equations for the corrections in the form
\begin{align}
\label{eq:DcorP22}
&\frac{d^2}{dx^2}D_x^{(1)} =D_x^{(1)}\pm 2\pi \lambda x \exp(\mp x), \nonumber \\
&\frac{d^2}{dx^2}D^{(1)}_y =D^{(1)}_y.
\end{align}
Corrections should vanish both at the infinity and near the source $D_{x,y}^{(1)}(x\to 0)=0$ and $D_{x,y}(\pm\infty)=0$, consequently 

\begin{align} \label{eq:swcor}
    &D_x^{(1)}=\mp\frac{ \pi \lambda }{2}x(1\pm x)\exp(\mp x), \nonumber \\
    &D_y^{(1)}=0.
\end{align}
Combining Eq.\eqref{eq:D01} with Eq.\eqref{eq:swcor} we finally arrive at

\begin{align}
\label{eq:swup}
   &D_x=\pm 2\pi \lambda \exp(\mp x)\mp \frac{g \pi \lambda }{2}x(1\pm x)\exp(\mp x), \\
   &D_y=0. \nonumber
\end{align}
Eq.\eqref{eq:swup} fully defines the shock electromagnetic wave produced by the infinitely long (in $y$) line that propagates along the $z$- axis.    

\section{Shape of the plasma bubble}

\subsection{Ballistic approximation}

As the first step in our considerations we neglect the effect of the plasma self-fields on the trajectories of the plasma electrons. This is a “ballistic” regime of plasma motion introduced in Ref.\cite{Stupakov:2016}; it assumes that the plasma electrons are moving with constant velocities.

We assume plasma electrons to be non-relativistic and $v_{z0}\approx1$, consequently for the upper half plane we get  
\begin{align}
\label{eq:trvel}
    &\frac{dx}{d\xi}\approx-2\pi \lambda \exp(-x_0)+\frac{g \pi \lambda }{2}x_0(1+x_0)\exp(-x_0).
\end{align}
Solution to the equations above gives electron trajectories
\begin{align}
\label{eq:trjf}
    x=&x_0-2\pi \lambda \xi \exp(-x_0)+ \nonumber\\ &\frac{g \pi \lambda \xi}{2}x_0(1+x_0)\exp(-x_0).
\end{align}

\begin{figure}[t]
	\centering
	\includegraphics[width=1.\linewidth]{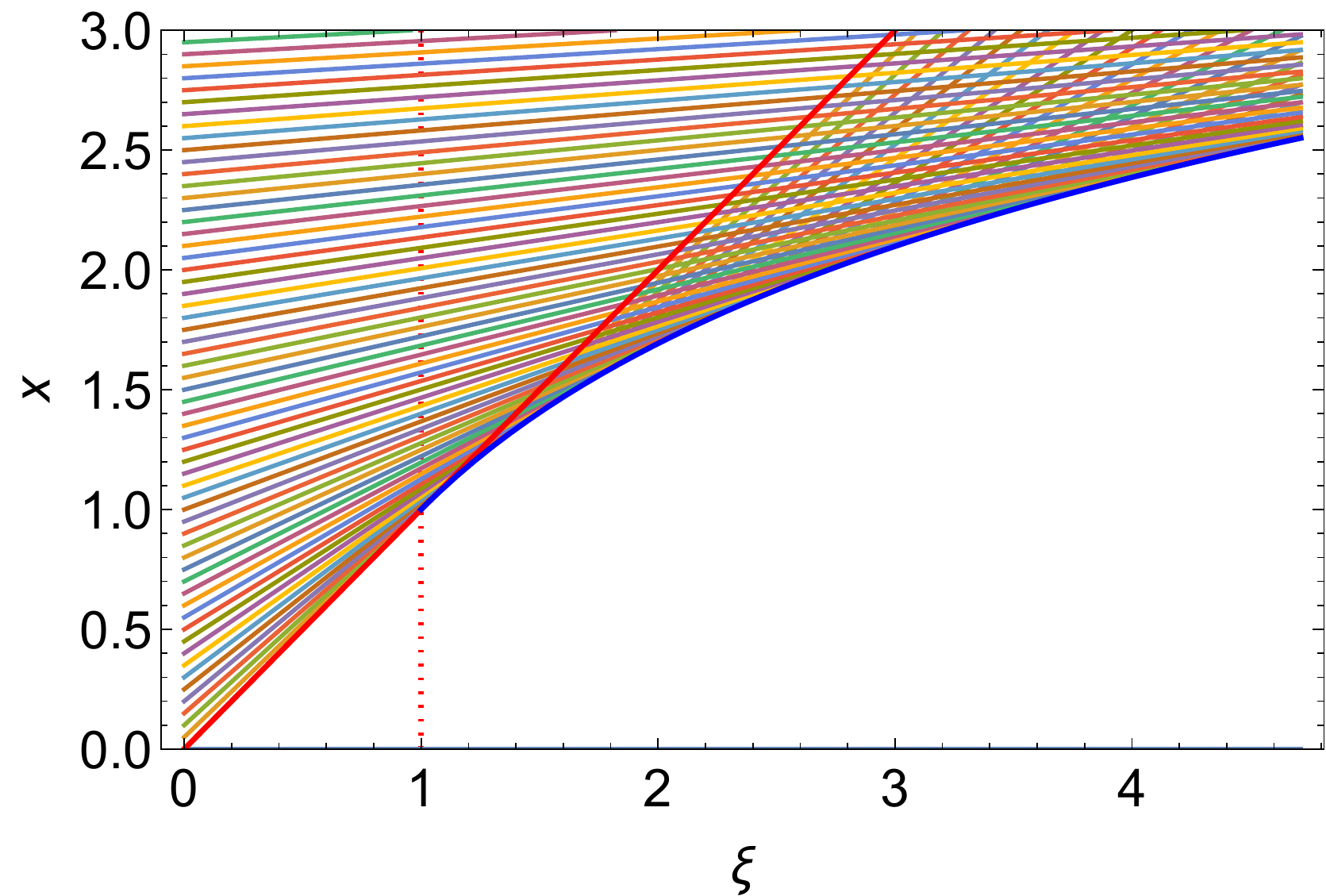}
	\caption{\label{fig:PSw} Plasma flow in the ballistic approximation for the case of $|\lambda|=1/2\pi$ and $g=0$. Red line denotes trajectory with $x_0=0$ Eq.\eqref{eq:p1}, and blue line denotes the envelope of the two stream flow given by Eq.\eqref{eq:p2}, dashed line shows switching point for the plasma flow.}
\end{figure}

Fist we consider the case of $g=0$ and plot electron trajectories in Fig.\Ref{fig:PSw} setting for the simplicity $\lambda=1/2\pi$. 

From Fig.\Ref{fig:PSw} we observer that plasma flow has two regimes: laminar flow, when trajectories do not cross and two stream flow. Switching point $\xi_{sw}$ could be found from the following considerations. We consider the upper half plane ($x>0$). Two stream flow appears when the most inner electron trajectory crosses the next closest trajectory. This condition could be expressed as
\begin{align}
  -2\pi \lambda \xi_{sw}=\delta x_0 -2\pi \lambda \xi_{sw} \exp(-\delta x_0).
\end{align}
Here $\delta x_0$ is the distance between starting points of the two trajectories. Solving for $\xi_{sw}$ and assuming $\lambda<0$ we get
\begin{align}
    \xi_{sw}=\frac{1}{2\pi |\lambda|} \frac{\delta x_0}{1-\exp(-\delta x_0)}.
\end{align}

Proceeding with the limit $\delta x_0 \to 0$ we finally have

\begin{align}
    \xi_{sw}=\frac{1}{2 \pi |\lambda|}.
\end{align}

Before the switching point bubble has a linear "triangular" shape and the boundary is simply defined by the most inner trajectory
\begin{align}
    x_b=2\pi |\lambda| \xi,~~~ \xi<\xi_{sw}.
\end{align}
Bubble boundary for the two stream flow (for $\xi \geq \xi_{sw}$) could be found as an envelope for the plasma electron trajectories from the condition
\begin{align}
\label{eq:p1}
    \frac{dx}{dx_0}=0,
\end{align}
and reads 
\begin{align}
\label{eq:p2}
    x_b=1+\mathrm{ln} \left(2\pi |\lambda| \xi \right),~~~ \xi\geq\xi_{sw}.
\end{align}
Next we account for the plasma gradient (now $g\neq 0$). Plasma flow still have two regimes and the switching point now should be found from the the condition 
\begin{align}
  2\pi |\lambda| \xi_{sw}&=\delta x_0 +2\pi |\lambda| \xi_{sw} \exp(-\delta x_0)-\nonumber \\ &\frac{g \pi |\lambda| \xi_{sw}}{2}\delta x_0(1+\delta x_0)\exp(-\delta x_0).
\end{align}
Solving for $\xi_{sw}$ gives
\begin{align}
    &\xi_{sw}= \nonumber \\
    &\frac{1}{2\pi |\lambda|} \frac{\delta x_0}{1-\exp(-\delta x_0)+\frac{g}{4}\delta x_0(1+\delta x_0)\exp(-\delta x_0)}.
\end{align}
Proceeding with the limit $\delta x_0 \to 0$ we have

\begin{align}
    \xi_{sw}=\frac{1}{2 \pi |\lambda|}.
\end{align}

We note that in the case of a small plasma gradient, the switching point remains at the same location as in the case of the uniform plasma and the first part of the plasma boundary remains the same and given by Eq.\eqref{eq:p1}. It is worth to mention that if $\lambda \ll 1$ then switching point $\xi_{sw}\to \infty$ and plasma flow is laminar. We note that in the case $g\neq 0$ boundary for the two stream part of the flow could not be expressed in a simple form as Eq.\eqref{eq:p1} becomes a transcendental equation. However, the flow could be easily found numerically.  

\subsection{Small charge regime \label{sec:Fl}}

Now we switch to a different approximation and do not neglect shielding anymore.
First we derive general expression for the force that acts on the plasma electrons. Form Eq.\eqref{eq:2dpp} with Eq.\eqref{eq:2dF} in the case of the non-relativistic plasma electrons ($v_z \ll 1$)  we get
\begin{align}
    \frac{dF_x}{dx}=n_i-n_e.
\end{align}
Integrating equation above we get
\begin{align}
\label{eq:Fx1}
    F_x(x)=\int\limits_{0}^x (n_i-n_e)d\tilde x+F_x(0).
\end{align}
At rest plasma is electrically neutral so the total charge of the electrons and ions should be the same in both lower ($x<0$) and upper ($x>0$) half planes. As far as the charge is conserved one may write $\int\limits_{0}^\infty (n_i-n_e)d\tilde x=0$, consequently from Eq.\eqref{eq:Fx1} we have $F_x(\infty)=F_x(0)$. Lorentz force $F_x$ have to vanish at the infinity, consequently we conclude that $F_x(0)=0$. With this observation one may rewrite Eq.\eqref{eq:Fx1} as      

\begin{align}
\label{eq:Fx2}
    F_x(x)=\int\limits_{0}^x (n_i-n_e)d\tilde x.
\end{align}
We notice that if the two stream regime does not develop then the shape of the bubble is always defined by the most inner trajectory of plasma electron that starts at $x_0=0$. Force that acts on this electron from the side of the plasma ions could be found from Eq.\eqref{eq:Fx2} with $n_e=0$ (inside the bubble there are no electrons). For the unperturbed case of $g=0$ we get familiar expression of the ion focusing force
\begin{align}
 F_x^i=x,~~~x\leq x_b.
\end{align}
We note that $F_x$ was defined as a force per unit charge, consequently the force that acts on the plasma electron is recovered by multiplying $F_x$ by the charge of the plasma electron. In our notations it is simply $-1$. 

Using initial conditions $x_0=0$ and $V_0=-2\pi \lambda$ one can arrive at the expression for the bubble shape in the form
\begin{align}\label{eq:bsp}
    x_b=2\pi |\lambda| \sin(\xi).
\end{align}
The range of validity for the formula above could be established from the condition of the crossing of the two closest trajectories and following  the same steps as in the ballistic approximation one can deduce equation for the switching point in the form 
\begin{align}
\label{eq:pp1}
    \sin(\xi_{sw})=\frac{1}{2\pi |\lambda|}.
\end{align}
For this equation to have no real solutions for the $\xi_{sw}$ the inequality $\sin(\xi_{sw})>1$ should hold, consequently 
\begin{align}
\label{eq:osrt}
    |\lambda|<\frac{1}{2\pi}.
\end{align}
Condition above sets the upper limit on the line charge density for the plasma flow to be laminar and described by Eq.\eqref{eq:bsp}.

\begin{figure}[t]
	\centering
	\includegraphics[width=1.\linewidth]{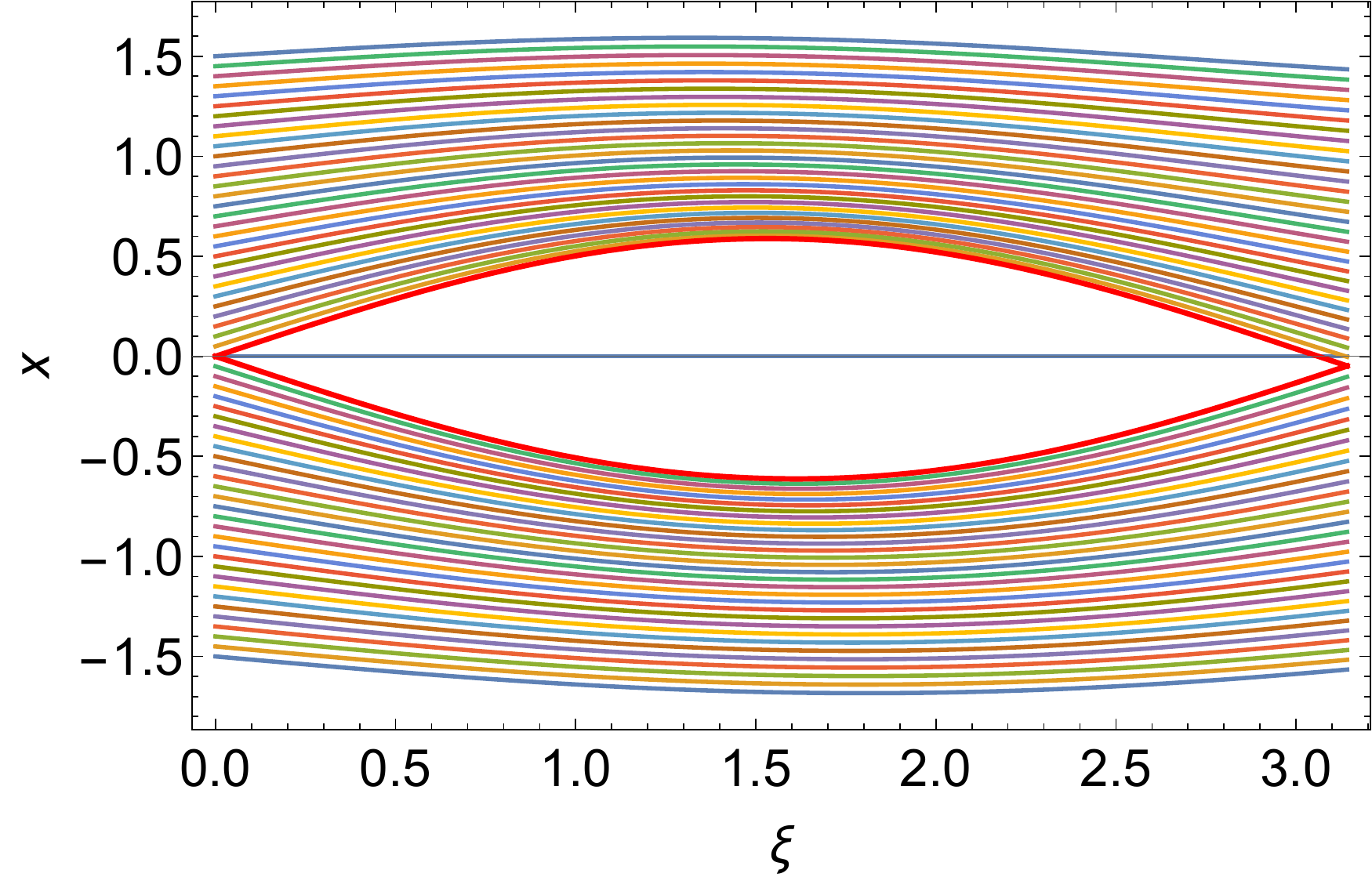}
	\caption{\label{Fig:PF} Left panel - plasma flow in the approximation of a small charge for the case of $\lambda=0.6/2\pi$ and $g=0.2$. Thick red lines denote bubble boundary.}
\end{figure}

Next we notice that if the condition \eqref{eq:osrt} significantly fulfilled then small perturbation to the ion and electron densities as well as to the ion force should not induce two stream regime and the plasma flow should remain laminar. Consequently, one may account for the ion density gradient 
\begin{align}
\label{eq:iond}
    n_i=1+g x,
\end{align}
with the help of Eq.\eqref{eq:Fx2} we get 
\begin{align}
\label{eq:ionF}
F_x^i=x+g\frac{x^2}{2},~~~x\leq x_b.    
\end{align}
As far as bubble boundary is given by the most inner electron trajectory equation for the bubble boundary reads
\begin{align}
\label{eq:bb_md}
    \frac{d^2x_b}{d \xi^2}=-x_b-g\frac{x_b^2}{2}.
\end{align}
with the initial conditions $x_b(0)=0$ and $x_b'(0)=2\pi \lambda$.
Equation \eqref{eq:bb_md} could be solved using perturbation method and solution up to the terms of the order $\mathcal{O}[ g^2]$ reads
\begin{align}
\label{eq:pp2}
    x_b=2\pi |\lambda| \sin( \xi)-\frac{8 \pi^2\lambda^2}{3}g\left[\sin (\frac{\xi}{2})\right]^4.
\end{align}
From Eq.\eqref{eq:pp1} we observe that in the case of the laminar flow and no plasma gradient bubble shape is universal and  scales linearly with the charge. However, in the case of the plasma gradient as follows from Eq.\eqref{eq:pp2} bubble shape always depends on the charge. Introducing normalized variable $x^n_b=x_b/(2\pi |\lambda|)$ we arrive at the final bubble shape in the form
\begin{align}
\label{eq:unv_b}
    x_b^n=\sin( \xi)-g\frac{4 \pi |\lambda|}{3}\left[\sin (\frac{\xi}{2})\right]^4.
\end{align}
We conclude that transverse plasma gradient results in "bending" of the bubble towards less dense plasma. Qualitatively this asymmetry is proportional to the product of the driver charge density and the strength of the plasma gradient.

\section{Plasma flow and plasma density}
%

Immediately behind the driver, at $\xi=0^+$, the plasma density $n_0$ is given by Eq.\eqref{14}. 

If we assume that electron trajectories are known then from  the continuity of the plasma flow we conclude that 
\begin{align}
n_e(x,y,\xi) dS = n_0(x_0,y_0) dS_0 
\end{align}
from which it follows that
\begin{align}
\label{eq:pldn}
n_e(x,y,\xi) = n_0(x_0,y_0) \frac{dS_0}{dS}. 
\end{align}
Accounting for the translation symmetry in $y$ we simply have  
\begin{align}
n_e(x,\xi) = n_0(x_0) \frac{dx_0}{dx}. 
\end{align}
or, substituting $n_0$ from Eq.\eqref{14}, we have
\begin{align}
\label{eq:eld2d}
   n_e(x,\xi) = (1+g x_0) \frac{dx_0}{dx}.  
\end{align}
We note that if plasma flow is laminar (electron trajectories do not cross) then force that is associated with the plasma electrons could be easily found with the help of Eq.\eqref{eq:Fx2} and with Eq.\eqref{eq:eld2d} reads
\begin{align}
\label{eq:trWe}
    F_x^e=-\int\limits_0^x (1+g x_0) \frac{dx_0}{dx}dx=-x_0(x)-g \frac{x_0(x)^2}{2}.
\end{align}
Interestingly this leads to a closed form solution for the electron trajectories as the equation of motion for each plasma electron with the initial conditions $x(0)=x_0$ and $x'(0)=-D_x(x_0)$ now reads
\begin{align}
\label{eq:gentrj}
    \frac{d^2x}{d \xi^2}=-x-g\frac{x^2}{2}+x_0+g \frac{x_0^2}{2}.
\end{align}
Equation above could be again solved using perturbation approach. 
\begin{align}
\label{eq:xper}
    x=x^{(0)}+g x^{(1)}+\mathcal{O} (g^2).
\end{align}
With Eq.\eqref{eq:swup} equation for the leading order $x^{(0)}$ reads
\begin{align}
    \frac{d^2x^{(0)}}{d \xi^2}&=-x^{(0)}+x_0, \nonumber\\
    x^{(0)}(0)&=x_0, \\
   \left. \frac{d x^{(0)}}{d\xi}\right|_{\xi=0}&=\pm 2 \pi |\lambda| \exp(\mp x_0). \nonumber
\end{align}
Here the upper sign is for the case $x>0$ and lower sign is for case $x<0$.
Solution to this equation reads
\begin{align}
\label{eq:xz}
    x^{(0)}=x_0\pm 2\pi |\lambda| \exp(\mp x_0) \sin (\xi).
\end{align}

\begin{figure}[t]
	\centering
	\includegraphics[width=0.9\linewidth]{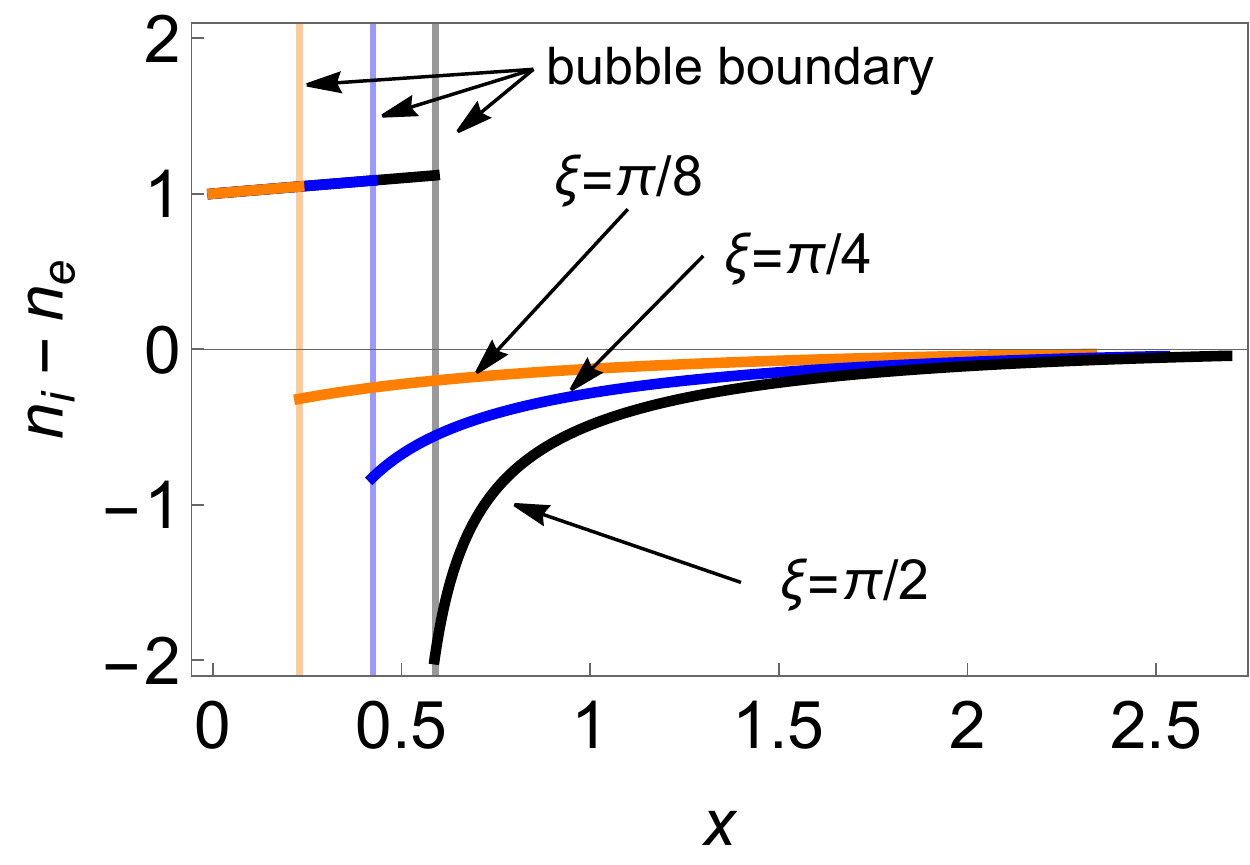}
	\caption{\label{Fig:Pld} Total plasma density for the case of $x>0$ and three different values of $\xi$ Inside the bubble electron density $n_e$ is equal to zero and the total density is essentially ion density given by Eq.\eqref{eq:iond}.$|\lambda|=0.6/2\pi$ and $g=0.2$.}
\end{figure}

Next to the leading order $x^{(1)}$ could be found by substituting Eq.\eqref{eq:xper} with Eq.\eqref{eq:swup} into the Eq.\eqref{eq:gentrj} and equating terms of the same order. After some algebra we have
\begin{align}
\label{eq:xse}
    \frac{d^2x^{(1)}}{d \xi^2}&=-x^{(1)}-\frac{\left(x^{(0)}\right)^2}{2}+\frac{x_0^2}{2}, \nonumber\\
    x^{(1)}(0)&=0, \\
   \left. \frac{d x^{(1)}}{d\xi}\right|_{\xi=0}&=\mp \frac{\pi |\lambda| }{2}x_0(1\pm x_0)\exp(\mp x_0). \nonumber
\end{align}
After substituting Eq.\eqref{eq:xz} into the Eq.\eqref{eq:xse} expression for the $x^{(1)}$ is found to be  
\begin{align}
\label{eq:x1fin}
    x^{(1)}=&-\frac{8 \pi^2 |\lambda|^2}{3}\exp(\mp 2x_0)\left[\sin (\frac{\xi}{2})\right]^4 \nonumber \\&\pm \pi |\lambda| x_0 \xi\exp(\mp x_0) \cos(\xi) \nonumber \\
    &\mp\frac{\pi |\lambda| x_0}{2} \exp(\mp x_0) \left[3\pm x_0\right]\sin(\xi).
\end{align}
Combining Eq.\eqref{eq:xz} and Eq.\eqref{eq:x1fin} with the help of the Eq.\eqref{eq:xper} we plot electron trajectories in Fig.\ref{Fig:PF} (left panel) for the case $g=0.2$ (moderately in-homogeneous plasma density) and $|\lambda|=0.6/2\pi$. As expected, the flow is laminar and modification to the bubble shape is very small (of the order $\sim \lambda^2 g$) as dictated by the Eq.\eqref{eq:pp2}.

To examine the total plasma density $n(x,\xi)=n_i(x,\xi)-n_e(x,\xi)$ we compare several cases ($n(x,\pi/8$), $n(x,\pi/4)$ and $n(x,\pi/2)$) in Fig.\ref{Fig:Pld}. The plot is produced with the help of the Eq.\eqref{eq:eld2d}. As far as the flow is laminar the solution to the Eq.\eqref{eq:xper} with Eq.\eqref{eq:xz} and Eq.\eqref{eq:x1fin} is unique with respect to $x_0$ for a given $x$ and could be found numerically. Consequently,
one may derive explicit expression for the plasma density in terms of $x_0$ and using this numerical solution compute $n(x_0(x),\xi)$. We do not provide final expressions as they are bulky but could be easily produced once needed.  

As could be seen from the Fig.\ref{Fig:Pld} in contrast to the cylindrical symmetric case (see Ref.\cite{Stupakov:2016}) plasma density does not have a singularity at the plasma boundary, however it has a jump that corresponds to the jump in the electron density. We observe as well, that as expected jump in the electron density increases towards the bubble wall and is maximal at the bubble boundary.

\begin{figure}[t]
	\centering
	\includegraphics[width=0.9\linewidth]{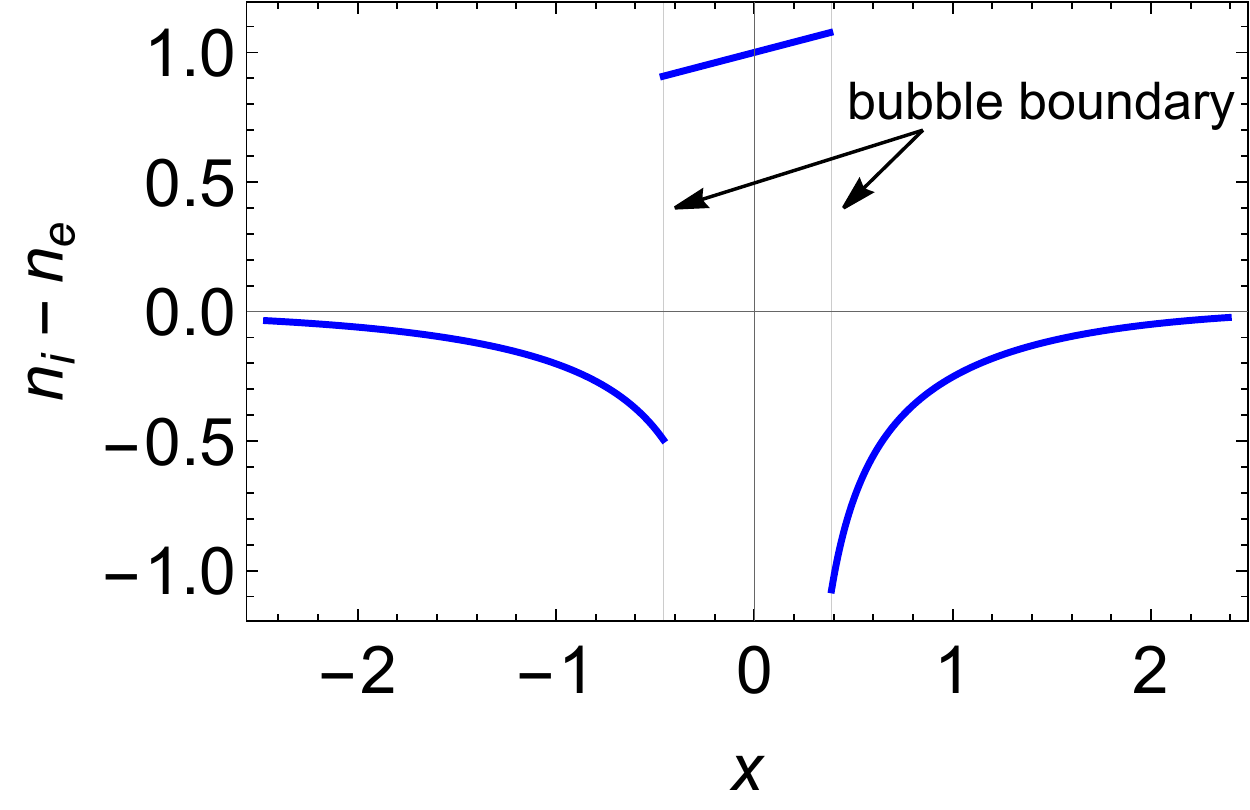}
	\caption{\label{Fig:Pld2} Total plasma density for the case of $\xi=3 \pi/4$, $|\lambda|=0.6/2\pi$ and $g=0.2$.}
\end{figure}

In Fig.\ref{Fig:Pld2} we plot plasma density close to the end of the bubble $\xi=3\pi/4$ where the witness beam is placed to maximize acceleration rate. One can observe that modest initial seed in plasma density gradient $g=0.2$ results in a sensible imbalance of the total plasma density at the bubble boundary. This in turn may result in the asymmetry of the witness self wake and consequently may affect the emittance of the witness beam.  

\section{Wakefield}

It was established in Sec.\ref{sec:Fl} that in the small charge regime a flat driver does not produce any transverse wake even in the case of the non-homogeneous plasma.
The force is purely focusing inside the bubble and is given by Eq.\eqref{eq:ionF}. 
Combining first and second formula in Eq.\eqref{eq:2dF} we get
\begin{align}
\label{eq:PWg}
    \frac{d E_z}{d x}=-\frac{d F_x}{d \xi},
\end{align}
The expression above is widely known as Panofsky-Wenzel theorem (see Ref.\cite{Panofsky:1956,Chao}) 
From Eq.\eqref{eq:PWg} we immediately conclude that $E_z$ is constant inside the bubble and depends on $\xi$ only. 
 
 Substitution of Eq.\eqref{eq:Fx1} into Eq.\eqref{eq:PWg} gives
 \begin{align}
\label{eq:PW}
    \frac{d E_z}{d x}=-\frac{d F^e_x}{d \xi},
\end{align}
 with $F_x^e$ given by Eq.\eqref{eq:trWe}. Expanding the $\xi$ derivative on the right hand side and integrating over $x$ we get for the $E_z$ inside the bubble
 
 \begin{align}
     E_z=\int\limits_{x_b}^\infty \frac{d x_0}{d x}\frac{dx}{d\xi}dx+g\int\limits_{x_b}^{\infty}x_0 \frac{dx_0}{dx}\frac{dx}{d\xi}dx. 
 \end{align}
 Interestingly both integrals in the expression above could be evaluated explicitly. Indeed, switching from integration over $x$ to integration over $x_0$ we get 
  \begin{align}
     E_z=\int\limits_{0}^\infty \frac{dx (x_0)}{d\xi}dx_0+g\int\limits_{0}^{\infty}x_0 \frac{dx (x_0)}{d\xi}dx_0. 
 \end{align}
 with $x(x_0)$ defined through Eq.\eqref{eq:xper} and $x^{(0)}$ and $x^{(1)}$ given by Eq.\eqref{eq:xz} and Eq.\eqref{eq:x1fin} respectively. 
 
 \begin{figure}[t]
	\centering
	\includegraphics[width=0.9\linewidth]{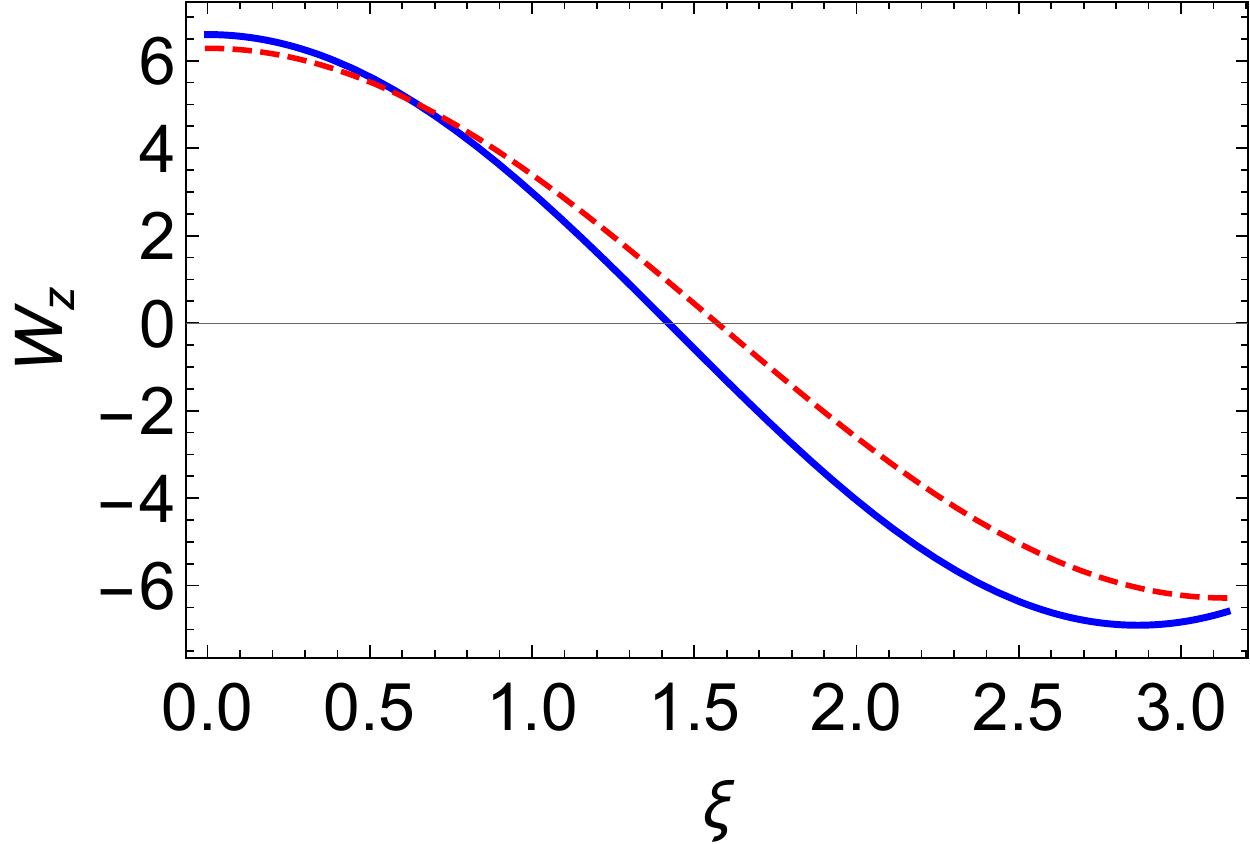}
	\caption{\label{Fig:Wz} Longitudinal wake potential per unit length $W_z=-E_z/\lambda$ for the case $|\lambda|=0.6/2\pi$, $g=0.2$ (blue line) and $g=0$ (red dashed line).}
\end{figure}
 
 After some algebra we finally get 
 \begin{align}
   E_z&=-2\pi \lambda \cos \left(\xi \right) \\&-g \pi  \lambda  \left[\frac{\cos (\xi)}{2}+\sin (\xi) \left[\frac{2 \pi  \lambda
   }{3}-\frac{2}{3} \pi  \lambda  \cos (\xi)-\xi\right]\right]  \nonumber.
 \end{align}
We observer that if the plasma is homogeneous than expression for the electric field is exactly the same as in the case of the linear plasma response despite the bubble formation. Local density gradient, however leads to the slight (proportional to the magnitude of the density gradient $g$) nonliterary in the wakefield (see Fig.\ref{Fig:Wz}).

We would like to reiterate two important conclusions. In the case of a laminar plasma flow an infinitely flat driver does not produce any transverse wake even if the local plasma density fluctuates from the equilibrium. In the laminar regime of the plasma flow in the flat case the longitudinal component of the electric field $E_z$ generated by the driver is constant inside the bubble and in the case of a homogeneous plasma has the same harmonic from as in the case of the linear plasma response. 

\section{Conclusion}

In this paper we have suggested a simple model of the flat bubble formation by the flat driver beam. Using suggested model and ballistic approximation introduced in Ref.\cite{Stupakov:2016} we have demonstrates that in the case of the flat drive two regimes (laminar and turbulent) could develop. The switching point where one type of the flow switches to another depends solely on the line charge density of the driver.       

We have investigated laminar plasma flow regime that is generate by the flat electron driver. It was demonstrated that a small perturbation to the plasma density results in "bending" of the bubble towards lower plasma gradient. Despite such a modification to the bubble shape and symmetry breaking, in the case of the non-relativistic plasma flow the wake generated by the flat driver does not have any deflecting components. Focusing force of the ion column has the same gradient as the initial gradient of the plasma density. It is worth mentioning that this gradient is independent of the longitudinal coordinate $\xi$ and consequently should not affect the emittance of the accelerated beam.

The model and results could be used for the analysis of the flat beam injection into the plasma cell that is an ongoing project at AWA facility \cite{Gerard}. 

One important observation that is worth reiterating is that in contrast to the approach of Ref.\cite{Stupakov:2016} the model considered in the present paper due to the 2D nature does not have a singularity neither at the position of the driver nor at the plasma boundary. This results in the formation of the transition region from the laminar flow to the two-stream turbulent flow.
In reality even in the case of the round driver the field is always finite at the driver position, consequently somewhat close to what is observer in the present mode should occur. Namely there should be always a transition of the laminar flow to the turbulent flow. In the case of the laminar flow the wake and its response to the external perturbation is expected to be different.

\begin{acknowledgments}
The author is grateful to G. Stupakov for fruitful discussions. The work was supported by the Government of the Russian Federation through the ITMO Fellowship and Professorship Program.
\end{acknowledgments}

\bibliographystyle{unsrt}
\bibliography{inhom_plasma}

\begin{thebibliography}{10}

\bibitem{GVoss}
G.A. Voss and T.~Weiland.
\newblock The wake field acceleration mechanism.
\newblock {\em DESY Report 82-074}, 1982.

\bibitem{Chen}
Pisin Chen, J.~M. Dawson, Robert~W. Huff, and T.~Katsouleas.
\newblock Acceleration of electrons by the interaction of a bunched electron
  beam with a plasma.
\newblock {\em Phys. Rev. Lett.}, 54:693--696, Feb 1985.

\bibitem{Rosen1}
J.~B. Rosenzweig.
\newblock Nonlinear plasma dynamics in the plasma wake-field accelerator.
\newblock {\em Phys. Rev. Lett.}, 58:555--558, Feb 1987.

\bibitem{Rosen2}
J.~B. Rosenzweig, B.~Breizman, T.~Katsouleas, and J.~J. Su.
\newblock Acceleration and focusing of electrons in two-dimensional nonlinear
  plasma wake fields.
\newblock {\em Phys. Rev. A}, 44:R6189--R6192, Nov 1991.

\bibitem{SNM3}
Spencer Gessner, Erik Adli, Weiming An, Sebastien Corde, Richard D'Arcy, Eric
  Esaray, Anna Grassellino, Bernhard Hidding, Mark Hogan, Ahmad~Fahim Habib,
  Axel Heubl, Chan Joshi, Wim Leemans, R.~Lehe, Carl Lindstr{\o}m, Michael
  Litos, Wei Lu, Warren Mori, Sergei Nagaitsev, Brendan O'Shea, Jens Osterhoff,
  Hasan Padamesee, Michael Peskin, Sam Posen, John Power, Tor Raubenheimer,
  James Rosenzweig, Marc Ross, Carl Schroeder, Paul Scherkl, Navid
  Vafaei-Najafabadi, Jean-Luc Vay, Glen White, and Vitaly Yakimenko.
\newblock Path towards a beam-driven plasma linear collider.
\newblock {\em {SNOWMASS}-21, LOI}, 2020.

\bibitem{LK}
L.K. Len.
\newblock Report of the doe advanced accelerator concepts research roadmap
  workshop.
\newblock {\em DOE, Gaithersburg, MD}, 2016.

\bibitem{alegro-2019-a}
ALEGRO collaboration.
\newblock Towards an advanced linear international collider.
\newblock {\em arXiv}, 1901.10370, 2019.

\bibitem{Adli}
Erik Adli.
\newblock Plasma wakefield linear colliders -- opportunities and challenges.
\newblock {\em Philosophical Transactions of the Royal Society A: Mathematical,
  Physical and Engineering Sciences}, 377(2151):20180419, 2019.

\bibitem{Sasha1}
Alex Murokh, Pietro Musumeci, Alexander Zholents, and Stephen Webb.
\newblock Towards a compact high efficiency fel for industrial applications.
\newblock In {\em OSA High-brightness Sources and Light-driven Interactions
  Congress 2020 (EUVXRAY, HILAS, MICS)}, page EF1A.3. Optica Publishing Group,
  2020.

\bibitem{Jamie1}
J~B Rosenzweig, N~Majernik, R~R Robles, G~Andonian, O~Camacho, A~Fukasawa,
  A~Kogar, G~Lawler, Jianwei Miao, P~Musumeci, B~Naranjo, Y~Sakai, R~Candler,
  B~Pound, C~Pellegrini, C~Emma, A~Halavanau, J~Hastings, Z~Li, M~Nasr,
  S~Tantawi, P.~Anisimov, B~Carlsten, F~Krawczyk, E~Simakov, L~Faillace,
  M~Ferrario, B~Spataro, S~Karkare, J~Maxson, Y~Ma, J~Wurtele, A~Murokh,
  A~Zholents, A~Cianchi, D~Cocco, and S~B van~der Geer.
\newblock An ultra-compact x-ray free-electron laser.
\newblock {\em New Journal of Physics}, 22(9):093067, sep 2020.

\bibitem{Barov}
N.~Barov, J.~B. Rosenzweig, M.~C. Thompson, and R.~B. Yoder.
\newblock Energy loss of a high-charge bunched electron beam in plasma:
  Analysis.
\newblock {\em Phys. Rev. ST Accel. Beams}, 7:061301, Jun 2004.

\bibitem{Lu}
W.~Lu, C.~Huang, M.~Zhou, W.~B. Mori, and T.~Katsouleas.
\newblock Nonlinear theory for relativistic plasma wakefields in the blowout
  regime.
\newblock {\em Phys. Rev. Lett.}, 96:165002, Apr 2006.

\bibitem{Stupakov:2016}
G.~Stupakov, B.~Breizman, V.~Khudik, and G.~Shvets.
\newblock Wake excited in plasma by an ultrarelativistic pointlike bunch.
\newblock {\em Phys. Rev. Accel. Beams}, 19:101302, Oct 2016.

\bibitem{Stupakov:2018}
G.~Stupakov.
\newblock Short-range wakefields generated in the blowout regime of
  plasma-wakefield acceleration.
\newblock {\em Phys. Rev. Accel. Beams}, 21:041301, Apr 2018.

\bibitem{Mori}
T.~N. Dalichaouch, X.~L. Xu, A.~Tableman, F.~Li, F.~S. Tsung, and W.~B. Mori.
\newblock A multi-sheath model for highly nonlinear plasma wakefields.
\newblock {\em Physics of Plasmas}, 28(6):063103, 2021.

\bibitem{QPIC}
C.~Huang, V.K. Decyk, C.~Ren, M.~Zhou, W.~Lu, W.B. Mori, J.H. Cooley, T.M.
  Antonsen, and T.~Katsouleas.
\newblock Quickpic: A highly efficient particle-in-cell code for modeling
  wakefield acceleration in plasmas.
\newblock {\em Journal of Computational Physics}, 217(2):658--679, 2006.

\bibitem{warp}
J-L Vay, D~P Grote, R~H Cohen, and A~Friedman.
\newblock Novel methods in the particle-in-cell accelerator code-framework
  warp.
\newblock {\em Computational Science {\&} Discovery}, 5(1):014019, dec 2012.

\bibitem{OSIRIS}
R.~A. Fonseca, L.~O. Silva, F.~S. Tsung, V.~K. Decyk, W.~Lu, C.~Ren, W.~B.
  Mori, S.~Deng, S.~Lee, T.~Katsouleas, and J.~C. Adam.
\newblock Osiris: A three-dimensional, fully relativistic particle in cell code
  for modeling plasma based accelerators.
\newblock In Peter M.~A. Sloot, Alfons~G. Hoekstra, C.~J.~Kenneth Tan, and
  Jack~J. Dongarra, editors, {\em Computational Science --- ICCS 2002}, pages
  342--351, Berlin, Heidelberg, 2002. Springer Berlin Heidelberg.

\bibitem{BL}
M.~Tzoufras, W.~Lu, F.~S. Tsung, C.~Huang, W.~B. Mori, T.~Katsouleas,
  J.~Vieira, R.~A. Fonseca, and L.~O. Silva.
\newblock Beam loading in the nonlinear regime of plasma-based acceleration.
\newblock {\em Phys. Rev. Lett.}, 101:145002, Sep 2008.

\bibitem{Timon1}
T.~J. Mehrling, R.~A. Fonseca, A.~Martinez de~la Ossa, and J.~Vieira.
\newblock Mitigation of the hose instability in plasma-wakefield accelerators.
\newblock {\em Phys. Rev. Lett.}, 118:174801, Apr 2017.

\bibitem{Timon2}
T.~J. Mehrling, C.~Benedetti, C.~B. Schroeder, E.~Esarey, and W.~P. Leemans.
\newblock Suppression of beam hosing in plasma accelerators with ion motion.
\newblock {\em Phys. Rev. Lett.}, 121:264802, Dec 2018.

\bibitem{Lehe}
R.~Lehe, C.~B. Schroeder, J.-L. Vay, E.~Esarey, and W.~P. Leemans.
\newblock Saturation of the hosing instability in quasilinear plasma
  accelerators.
\newblock {\em Phys. Rev. Lett.}, 119:244801, Dec 2017.

\bibitem{Carl}
Carl~A. Lindstr\o{}m.
\newblock Staging of plasma-wakefield accelerators.
\newblock {\em Phys. Rev. Accel. Beams}, 24:014801, Jan 2021.

\bibitem{Jing}
Chunguang Jing.
\newblock Dielectric wakefield accelerators.
\newblock {\em Reviews of Accelerator Science and Technology}, 09:127--149,
  2016.

\bibitem{Sasha2}
A.~Siy, N.~Behdad, J.~Booske, M.~Fedurin, W.~Jansma, K.~Kusche, S.~Lee,
  G.~Mouravieff, A.~Nassiri, S.~Oliphant, S.~Sorsher, K.~Suthar,
  E.~Trakhtenberg, G.~Waldschmidt, and A.~Zholents.
\newblock Fabrication and testing of corrugated waveguides for a collinear
  wakefield accelerator.
\newblock {\em Phys. Rev. Accel. Beams}, 25:021302, Feb 2022.

\bibitem{Sasha3}
A.~Zholents, S.~Baturin, S.~Doran, W.~Jansma, M.~Kasa, R.~Kustom, A.~Nassiri,
  J.~Power, K.~Suthar, E.~Trakhtenberg, I.~Vasserman, G.~Waldschmidt, and
  J.~Xu.
\newblock A compact wakefield accelerator for a high repetition rate multi user
  x-ray free-electron laser facility.
\newblock In {\em High-Brightness Sources and Light-driven Interactions}, page
  EW3B.1. Optica Publishing Group, 2018.

\bibitem{Hosing1}
David~H. Whittum, William~M. Sharp, Simon~S. Yu, Martin Lampe, and Glenn Joyce.
\newblock Electron-hose instability in the ion-focused regime.
\newblock {\em Phys. Rev. Lett.}, 67:991--994, Aug 1991.

\bibitem{BBU2}
C.~Li, W.~Gai, C.~Jing, J.~G. Power, C.~X. Tang, and A.~Zholents.
\newblock High gradient limits due to single bunch beam breakup in a collinear
  dielectric wakefield accelerator.
\newblock {\em Phys. Rev. ST Accel. Beams}, 17:091302, Sep 2014.

\bibitem{BBU3}
S.~S. Baturin and A.~Zholents.
\newblock Stability condition for the drive bunch in a collinear wakefield
  accelerator.
\newblock {\em Phys. Rev. Accel. Beams}, 21:031301, Mar 2018.

\bibitem{Piot1}
P.~Piot, Y.-E Sun, and K.-J. Kim.
\newblock Photoinjector generation of a flat electron beam with transverse
  emittance ratio of 100.
\newblock {\em Phys. Rev. ST Accel. Beams}, 9:031001, Mar 2006.

\bibitem{Piot2}
J.~Zhu, P.~Piot, D.~Mihalcea, and C.~R. Prokop.
\newblock Formation of compressed flat electron beams with high
  transverse-emittance ratios.
\newblock {\em Phys. Rev. ST Accel. Beams}, 17:084401, Aug 2014.

\bibitem{Piot2c}
A.~Halavanau, J.~Hyun, D.~Mihalcea, P.~Piot, T.~Sen, and J.C.T. Thangaraj.
\newblock {M}agnetized and {F}lat {B}eam {E}xperiment at {FAST}.
\newblock In {\em Proc. of International Particle Accelerator Conference
  (IPAC'17), Copenhagen, Denmark, 14-19 May, 2017}, pages 3876--3879. JACoW,
  2017.

\bibitem{Baturin:dDrive}
S.~S. Baturin, G.~Andonian, and J.~B. Rosenzweig.
\newblock Analytical treatment of the wakefields driven by transversely shaped
  beams in a planar slow-wave structure.
\newblock {\em Phys. Rev. Accel. Beams}, 21:121302, Dec 2018.

\bibitem{FLb}
A.~Tremaine, J.~Rosenzweig, and P.~Schoessow.
\newblock Electromagnetic wake fields and beam stability in slab-symmetric
  dielectric structures.
\newblock {\em Phys. Rev. E}, 56:7204--7216, Dec 1997.

\bibitem{Brendan:2020}
Brendan~D. O'Shea, Gerard Andonian, S.~S. Baturin, Christine~I. Clarke, P.~D.
  Hoang, Mark~J. Hogan, Brian Naranjo, Oliver~B. Williams, Vitaly Yakimenko,
  and James~B. Rosenzweig.
\newblock Suppression of deflecting forces in planar-symmetric dielectric
  wakefield accelerating structures with elliptical bunches.
\newblock {\em Phys. Rev. Lett.}, 124:104801, Mar 2020.

\bibitem{Joshi}
Shiyu Zhou, Jianfei Hua, Weiming An, Warren~B. Mori, Chan Joshi, Jie Gao, and
  Wei Lu.
\newblock High efficiency uniform wakefield acceleration of a positron beam
  using stable asymmetric mode in a hollow channel plasma.
\newblock {\em Phys. Rev. Lett.}, 127:174801, Oct 2021.

\bibitem{Gessner2}
Spencer Gessner, Erik Adli, James~M. Allen, Weiming An, Christine~I. Clarke,
  Chris~E. Clayton, Sebastien Corde, J.~P. Delahaye, Joel Frederico, Selina~Z.
  Green, Carsten Hast, Mark~J. Hogan, Chan Joshi, Carl~A. Lindstr{\o}m, Nate
  Lipkowitz, Michael Litos, Wei Lu, Kenneth~A. Marsh, Warren~B. Mori, Brendan
  O'Shea, Navid Vafaei-Najafabadi, Dieter Walz, Vitaly Yakimenko, and Gerald
  Yocky.
\newblock Demonstration of a positron beam-driven hollow channel plasma
  wakefield accelerator.
\newblock {\em Nat. Comm.}, 7(1):11785, 2016.

\bibitem{Gessner1}
Spencer~J. Gessner.
\newblock {\em Demonstration of the hollow channel plasma wakefield
  accelerator}.
\newblock PhD thesis, Stanford University, 9 2016.

\bibitem{Gerard}
P.~Manwani, H.S. Ancelin, G.~Andonian, G.~Ha, J.G. Power, J.B. Rosenzweig, and
  M.~Yadav.
\newblock {Asymmetric Beam Driven Plasma Wakefields at the AWA}.
\newblock In {\em Proc. IPAC'21}, number~12 in International Particle
  Accelerator Conference, pages 1732--1735. JACoW Publishing, Geneva,
  Switzerland, 08 2021.
\newblock https://doi.org/10.18429/JACoW-IPAC2021-TUPAB147.

\bibitem{Panofsky:1956}
W.~K.~H. Panofsky and W.~A. Wenzel.
\newblock Some considerations concerning the transverse deflection of charged
  particles in radio‐frequency fields.
\newblock {\em Review of Scientific Instruments}, 27(11):967--967, 1956.

\bibitem{Chao}
Alex Chao.
\newblock {\em Physics of Collective Beam Instabilities in High Energy
  Accelerators}.
\newblock Wiley and Sons, New York, 1993.

\end{thebibliography}

\end{document}